\documentclass[sigconf,nonacm]{acmart}

\newcommand{\etal}{et~al.\xspace}

\newcommand{\eg}{e.g.\xspace}
\newcommand{\rsq}{R\textsuperscript{2}\xspace}

\newcommand{\rqone}{%
    How prevalent is systemic flakiness?\xspace
}

\newcommand{\rqtwo}{%
    How well can machine learning models predict systemic flakiness using static 
    test case distance measures?\xspace
}

\newcommand{\rqthree}{%
    What causes systemic flakiness?\xspace
}

\newcommand{\conclusionbox}[2]{%
    \begin{mdframed}[
        skipabove=3pt,
        skipbelow=3pt,
        innertopmargin=3pt,
        innerleftmargin=3pt,
        innerrightmargin=3pt,
        innerbottommargin=3pt
      ]
        \noindent {\bf Conclusion for #1.} #2
    \end{mdframed}
}

\acmConference[EASE 2025]{%
    The 29th International Conference on Evaluation and Assessment in Software 
    Engineering
}{%
    17--20 June, 2025
}{%
    Istanbul, Türkiye
}

\title{%
    Systemic Flakiness: An Empirical Analysis of Co-Occurring Flaky Test 
    Failures
}

\author{Owain Parry}
\affiliation{\institution{University of Sheffield}\country{UK}}
\author{Gregory M. Kapfhammer}
\affiliation{\institution{Allegheny College}\country{USA}}
\author{Michael Hilton}
\affiliation{\institution{Carnegie Mellon University}\country{USA}}
\author{Phil McMinn}
\affiliation{\institution{University of Sheffield}\country{UK}}

\begin{abstract}
    {\it Flaky tests} produce inconsistent outcomes without code changes, 
    creating major challenges for software developers.
    An industrial case study reported that developers spend 1.28\% of their 
    time repairing flaky tests at a monthly cost of \$2,250.
    We discovered that flaky tests often exist in clusters, with 
    co-occurring failures that share the same root causes, which we call 
    {\it systemic flakiness}.
    This suggests that developers can reduce repair costs by addressing 
    shared root causes, enabling them to fix multiple flaky tests at once 
    rather than tackling them individually.
    This study represents an inflection point by challenging the deep-seated 
    assumption that flaky test failures are isolated occurrences.
    We used an established dataset of 10,000 test suite runs from 24 Java 
    projects on GitHub, spanning domains from data orchestration to job 
    scheduling.
    It contains 810 flaky tests, which we levered to perform a mixed-method 
    empirical analysis of co-occurring flaky test failures.

    Systemic flakiness is significant and widespread.
    We performed agglomerative clustering of flaky tests based on their 
    failure co-occurrence, finding that 75\% of flaky tests across all 
    projects belong to a cluster, with a mean cluster size of 13.5 flaky 
    tests.
    Instead of requiring 10,000 test suite runs to identify systemic 
    flakiness, we demonstrated a lightweight alternative by training machine 
    learning models based on static test case distance measures.
    Through manual inspection of stack traces, conducted independently by 
    four authors and resolved through negotiated agreement, we identified 
    intermittent networking issues and instabilities in external 
    dependencies as the predominant causes of systemic flakiness.
\end{abstract}

\ccsdesc[500]{Software and its engineering~Software testing and debugging}

\keywords{%
    Software Testing, 
    Flaky Tests, 
    Systemic Flakiness.
}

\thanks{%
    Owain and Phil are supported by the EPSRC grant "Test FLARE" (EP/X024539/1).
}

\usepackage{booktabs}
\usepackage{graphicx}
\usepackage{xspace}
\usepackage{balance}
\usepackage{tikz}
\usepackage{pgfplots}
\usepackage{amsmath}
\usepackage{colortbl}
\usepackage{mdframed}
\usepackage{xurl}
\usepackage{subcaption}
\usepackage{listings}
\usepackage{array}
\usepackage{xcolor}

\begin{document}
    \maketitle
    \section{Introduction}
\label{sec:introduction}

Software developers rely on test cases to identify bugs~\cite{Kapfhammer2004}.
However, when test results are unreliable, they lose their value as informative 
signals and developers may deem them untrustworthy~\cite{Gruber2022}.
Practitioners in software testing refer to such unreliable signals as {\it flaky 
tests}~\cite{Parry2021}.
While definitions vary~\cite{Parry2022b}, a flaky test is generally understood 
as a test case that can pass or fail unpredictably without changes to its code 
or the code under test.
Flaky tests may arise from concurrency bugs, timing issues, or dependencies on 
external systems like networks and filesystems~\cite{Luo2014b,Vahabzadeh2015,
Eck2019,Lam2020c,Gruber2021,Hashemi2022}.

Recent developer surveys and interviews underscore the significant challenges 
posed by flaky tests~\cite{Ahmad2021b,Gruber2022,Habchi2022}.
In one survey, 56\% of respondents reported encountering flaky tests on a 
monthly, weekly, or even daily basis~\cite{Parry2022b}.
Respondents also strongly agreed that flaky tests disrupt continuous 
integration~\cite{Hilton2017}, reduce productivity, and hinder testing 
efficiency.
The impact of flaky tests is felt across the industry, from large companies like 
Google and Microsoft~\cite{Memon2017,Lam2019b}, to open-source development 
communities~\cite{Eck2019,Durieux2020}.
An industrial case study on the cost of flaky tests in continuous integration
found that developers spend up to 1.28\% of their time repairing flaky tests, 
amounting to a monthly cost of \$2,250~\cite{Leinen2024}.

In this study, we found that flaky tests frequently exist in clusters, with 
failures that co-occur during the same test suite runs and share the same root 
causes.
We call this phenomenon {\it systemic flakiness}.
It implies that developers can reduce the cost of repairing flaky tests by 
targeting shared root causes, allowing them to simultaneously fix numerous flaky 
tests instead of addressing them in isolation.

Surprisingly, systemic flakiness has been neglected in prior research, making 
this study a key inflection point.
Previous studies have relied on simulated failures to assess the impact of flaky 
tests on milestone techniques in software engineering research, such as fault 
localization, mutation testing, and automated program repair~\cite{Vancsics2020,
Cordy2022}.
However, by failing to account for co-occurring flaky test failures, these 
simulations misrepresent real-world flakiness.
Similarly, prior studies categorized the causes of individual flaky tests 
without considering systemic flakiness~\cite{Luo2014b,Vahabzadeh2015,Eck2019,
Lam2020c,Gruber2021,Hashemi2022}.
These studies may have inadvertently reported a skewed distribution of causes 
and should be reexamined in light of this study.

We revisited an existing dataset of 10,000 test suite runs from 24 diverse 
open-source Java projects that has seen extensive use in previous 
studies~\cite{Alshammari2021,Fatima2022,Alshammari2024,Wang2024}.
This dataset contains 810 examples of flaky tests and required over five years 
of computation time to produce~\cite{Alshammari2021}.
We leveraged it to perform a mixed-method empirical analysis.
This was necessary because no single analysis method can fully capture the 
complexity of systemic flakiness.

This study combines the following methods: \\
{\bf Method 1: Clustering.} To analyze the prevalence of systemic flakiness, we 
performed agglomerative clustering based on the Jaccard distance between the 
sets of test suite runs in which flaky tests fail. \\
{\bf Method 2: Prediction.} To investigate the feasibility of a lightweight 
alternative to performing 10,000 test suite runs to identify systemic flakiness, 
we trained machine learning models to predict the Jaccard distance between pairs 
of flaky tests using test case distance measures based on static 
analysis~\cite{Elgendy2024,Elgendy2025}. \\
{\bf Method 3: Manual Inspection.} To identify systemic flakiness causes, we 
conducted a qualitative analysis of stack traces and error messages combining 
manual inspection and negotiated agreement.

This study answers the following research questions: \\
{\bf RQ1: \rqone} Across all projects, 75\% of flaky tests fail as part of a 
cluster, with a mean cluster size of 13.5 flaky tests spanning multiple test 
classes. \\
{\bf RQ2: \rqtwo} The prediction accuracy varies across projects, with the best 
model achieving a mean coefficient of determination of 0.74 when predicting the 
Jaccard distance between pairs of flaky tests. \\
{\bf RQ3: \rqthree} The main causes are networking issues and instabilities in 
external dependencies.

This study makes the following main contributions: \\
{\bf Contribution 1: Systemic Perspective.} This study introduces a systemic 
perspective on flaky tests, highlighting the significance of failure 
co-occurrence, which has been overlooked in prior research. \\
{\bf Contribution 2: Empirical Analysis.} This study quantitatively analyzes 
systemic flakiness using clustering techniques and qualitatively investigates 
the causes.
All the data and results from the empirical study are publicly available in 
our replication package~\cite{ReplicationPackage}. \\
{\bf Contribution 3: Practical Implications.} This study demonstrates the 
feasibility of leveraging static test case distance measures and machine 
learning models to predict systemic flakiness, offering an alternative to 
exhaustive test suite executions.

    \section{Methodology}
\label{sec:methodology}

In this section, we describe our methodology for answering our three research
questions regarding systemic flakiness.
We developed Python scripts to automate the key aspects of our analysis.
These are available in our replication package~\cite{ReplicationPackage}.

\subsection{Dataset}
\label{sec:methodology:dataset}

We used the dataset created by Alshammari \etal for their evaluation of 
FlakeFlagger, a machine learning-based technique for automatically detecting 
flaky tests~\cite{Alshammari2021}.
They selected 24 Java projects from GitHub that had been used in prior flaky 
test studies~\cite{Bell2018,Lam2019}.
The projects span a variety of domains from data orchestration to job 
scheduling.
They ran each projects' test suite 10,000 times, requiring over 5 years of 
computation time.
To do so, they created a queue of jobs, where every job represented a single 
test suite run (running {\tt mvn install}).
After each job, they archived all log files, removed all temporary files, and
rebooted the machine before proceeding to the next job in the queue.
They argued that this approach provided reasonable isolation between test suite 
runs and simulated how a real build server might compile and test a project.

This dataset contains a JUnit test report for every run of each project's test 
suite that includes the outcome (pass/fail) of every test case and the full 
stack trace and error message for test cases that failed.
It contains 810 examples of flaky tests (test cases with an inconsistent 
outcome) among 22 projects\footnote{%
    In their original paper, Alshammari \etal reported 811 flaky tests among 23
    projects~\cite{Alshammari2021}.
    However, one of the flaky tests did not appear to have any associated stack 
    traces or error messages, so we excluded it from this study.
}.
It also contains the source code of each test case in textual and tokenized 
form.
We selected this dataset due to the vast number of test suite runs, which is 
very useful for reliably analyzing co-occurring flaky test failures.
The available stack traces and source code are also useful for answering 
{\bf RQ2} and {\bf RQ3}.

\subsection{Methodology for RQ1 (Prevalence)}
\label{sec:methodology:rqone}

Our scripts performed {\it agglomerative clustering} of the flaky tests in the 
projects of the FlakeFlagger dataset using the SciPy library~\cite{Scipy2025}.
Agglomerative clustering is a type of clustering technique that starts by 
treating each data point as its own cluster and then iteratively merges the two 
closest clusters until all data points are merged into a single 
cluster~\cite{Ackermann2014}.
The result of agglomerative clustering is a hierarchy of clusters from which a
{\it concrete clustering} (an assignment of each data point to a specific 
cluster) can be extracted by specifying a distance threshold.
We selected a clustering approach because it allows us to identify specific
instances of systemic flakiness in a project by grouping flaky tests based on 
their failure co-occurrence.
We selected agglomerative clustering specifically because it is does not require 
us to prespecify the number of clusters in a project, which we could not 
possibly have known a priori.

Agglomerative clustering requires a distance metric to compare two data points.
In the context of this study, a data point is a flaky test and the distance 
metric needs to capture the extent to which the failures of two flaky tests 
co-occur.
In the FlakeFlagger dataset, each of the 10,000 runs of a project's test suite
is associated with a unique numerical ID.
As the distance metric to compare two flaky tests, we selected the {\it Jaccard 
distance} between the two sets of run IDs in which they fail.
The Jaccard distance $J$ between two sets $A$ and $B$ is a measure of how 
dissimilar they are.
It is defined as:
\begin{equation}
    \label{equ:jaccard}
    J(A, B) = 1 - \frac{|A \cap B|}{|A \cup B|}
\end{equation}

For example, suppose one flaky test fails during three test suite runs with IDs
$\{52, 901, 5810\}$, and another fails during five test suite runs with IDs
$\{52, 901, 1119, 5810, 9402\}$.
The length of the union of these two sets is 5 and they have 3 failing run IDs 
in common, so their Jaccard distance is $1 - 3 / 5 = 0.4$.
In this context, a Jaccard distance of 1 between two flaky tests indicates that 
there were no test suite runs where they both failed, and 0 indicates that they 
always failed during the same test suite runs.
In the latter case, this would imply that the two flaky tests likely share the 
same root cause and are therefore a manifestation of systemic flakiness.
We selected the Jaccard distance because it has a straightforward interpretation
and is robust when comparing two flaky tests with vastly different numbers of 
failing runs.
This is important for this study because we are primarily interested in failure
co-occurrence as opposed to the failure rates of individual flaky tests.

Our scripts automatically identified the distance threshold that extracts the 
concrete clustering with the greatest mean {\it silhouette score} for each 
project~\cite{Shahapure2020}.
For a single flaky test, the silhouette score $s$ evaluates how well it fits 
within its assigned cluster compared to other clusters.
It considers the average Jaccard distance $a$ to the other flaky tests in its 
assigned cluster and the average distance $b$ to the flaky tests in the nearest 
other cluster.
It is defined as:
\begin{equation}
    \label{equ:silhouette}
    s = \frac{b - a}{max(a, b)}
\end{equation}

The mean silhouette score over all the flaky tests in a project evaluates the
overall quality of the concrete clustering.
It ranges from 1 to -1 where scores above 0.7 indicate strong clusters 
(well-defined and compact), scores above 0.5 indicate reasonable clusters, and
scores above 0.25 indicate weak clusters~\cite{Kaufman1990}.
We assumed that there was no systemic flakiness in a project if there was no 
possible concrete clustering with a mean silhouette score of at least 0.6, which 
indicates clusters that are more than merely acceptable without being overly 
restrictive.
We also did not consider clusters containing only a single flaky test (known as 
{\it singleton clusters}) as evidence of systemic flakiness because they do not 
identify any failure co-occurrence by definition.

We implemented this automated approach of identifying a separate distance 
threshold for each project for several reasons.
Firstly, automatically selecting the distance threshold based on the mean 
silhouette score removes subjective bias and ensures that the clustering process 
is consistent and reproducible across projects.
Secondly, setting a minimum mean silhouette score of 0.6 ensures that clusters 
are reasonably well-defined and interpretable.
This avoids the risk of identifying spurious instances of systemic flakiness 
that may undermine the reliability of our analysis.
Finally, identifying a distance threshold for each project individually 
accommodates potential differences in flaky test behaviors, which may depend on 
the project's size, purpose, or complexity.

\subsection{Methodology for RQ2 (Prediction)}
\label{sec:methodology:rqtwo}

Detecting flaky tests without performing test suite runs has been extensively 
studied in prior work~\cite{Pinto2020,Bertolino2021,Camara2021,Pontillo2022,
Fatima2022,Qin2022}.
Techniques to do so typically make use of machine learning models trained on 
features based on static analysis of test case code, such as specific token 
occurrences and complexity metrics.
Building on this foundation, we set out to evaluate how effectively machine 
learning models can predict systemic flakiness based on static test case 
distance measures between pairs of flaky tests.
Such distance measures have been applied in prior work to assess test suite 
diversity~\cite{Elgendy2024,Elgendy2025}.
We selected them as features based on the intuition that the more similar two 
flaky tests are, the more likely their failures will co-occur.

Our scripts trained and evaluated three tree-based ensemble models in the 
context of a regression task and a classification task using the scikit-learn 
library~\cite{ScikitLearn2025}.
They did this for each project in the FlakeFlagger dataset with at least 10 
flaky tests and one cluster.
The three models we selected were extra trees, gradient boosting, and random 
forest~\cite{Breiman2001,Friedman2001,Geurts2006}.
We selected these models because they capture complex nonlinear relationships, 
perform well even with moderate amounts of training data, and are robust to 
overfitting thanks to their ensemble nature~\cite{Uddin2024}.
For each model, we used the default hyperparameters in scikit-learn.
The number of sub-models in the ensemble, arguably the most important 
hyperparameter, has a default value of 100 in each case.
For each pair of flaky tests in a project, the regression task is to predict the 
Jaccard distance between the two sets of failing run IDs and the classification
task is to predict whether they belong to the same cluster.

For the regression task, our scripts evaluated the performance of each model by 
calculating the mean {\it coefficient of determination} (\rsq) following 5-fold 
cross validation over the flaky tests in a given project.
This is a standard measure for regression tasks that indicates the amount of 
variance in the target variable (the Jaccard distance) that is explained by the 
model.
It ranges from 1 to negative infinity, where a value of 1 indicates perfect 
predictions.
A model that ignores the input features and always predicts the mean value of 
the target variable would score 0.
For the classification task, our scripts evaluated the performance of each model 
by calculating the mean {\it Matthews correlation coefficient} (MCC) following 
5-fold {\it stratified} cross validation.
This is a reliable metric for binary classification tasks that is robust to 
label imbalance~\cite{Chicco2020}.
It ranges from 1 to -1, where a value of 1 indicates perfect predictions.
A model that always predicts the most common label would score 0.

As features, we selected character-based and set-based distance measures applied 
to the names and source code of pairs of flaky tests.
The character-based measures we selected were the Levenshtein distance, the 
Damerau-Levenshtein distance, the Jaro distance, and the Jaro-Winkler distance.
We also used the normalized variants of the Levenshtein and Damerau-Levenshtein 
distances.
The set-based measures we selected were the Jaccard distance, the Dice distance, 
and the overlap distance (one minus the overlap coefficient).
To calculate the set-based distance measures, our scripts split the names of 
each flaky test into a set of tokens.
(The FlakeFlagger dataset already contains the source code of every test case in
tokenized form.)
A fully qualified test case name in Java contains the package name, class name,
and method name.
For example: {\tt package.name.ClassName\#methodName}.
Our scripts split these into a set of tokens by breaking up the components and 
splitting the class names and method names based on their capitalization.
Applied to the previous example, this would result in the set of unique tokens: 
{\tt package}, {\tt name}, {\tt class}, and {\tt method}.

We included an additional feature that is a normalized distance measure between 
two test case names based on their hierarchical structure that we call the 
{\it hierarchy distance}.
It is calculated by first extracting the {\it path} of the two test case names
by discarding the method name and splitting the remainder into its components.
For example, the path of the test case name {\tt foo.bar.Baz\#qux} would be the
sequence {\tt foo}, {\tt bar}, and {\tt Baz}.
Where $n$ is the length of the longer of the two paths and $i$ is the first 
index where the two paths differ following a zero-based indexing scheme, $n$
if they are identical, or the length of the shorter of the two paths if one is
a prefix of the other, the hierarchy distance is $1 - i / n$.
A value of 1 indicates that the two paths are identical and a value of 0 
indicates they they differ in their first component.
It captures the distance between two test cases within the hierarchical 
structure of the project test code.

To get a sense of which static test case distance measures are the most 
important features for systemic flakiness prediction, our scripts calculated 
{\it SHAP values} with respect to the regression task and the model with the 
greatest coefficient of determination~\cite{Lundberg2020}.
For a given distance measure and pair of flaky tests, the corresponding SHAP 
value captures the contribution of that distance measure towards the model's 
prediction of the Jaccard distance between the two sets of failing run IDs.
The sum of the SHAP values over each distance measure for a given pair of flaky 
tests is equal to the model's prediction of their Jaccard distance.
For each project, our scripts ranked the distance measures in terms of their 
mean absolute SHAP value over every pair of flaky tests for a general overview 
of which were the most impactful.

\subsection{Methodology for RQ3 (Causes)}
\label{sec:methodology:rqthree}

Four authors of this paper engaged in manual inspection of each cluster that our 
scripts identified for {\bf RQ1}.
We randomly allocated two inspecting authors to every project in the dataset 
with at least one cluster.
Each inspector then independently answered a series of questions about every 
cluster in their allocated projects in a random order.
One of the questions was ``What are the root causes of the flaky test failures 
in this cluster?'', which directly addresses {\bf RQ3}.
Another question was ``What actions could a developer take to repair or mitigate 
the flaky tests in this cluster?''.
Inspectors were also given the opportunity to offer any miscellaneous insights.

To enable the inspectors to answer these questions for a given cluster, they
were able to review the source code of its member flaky tests.
Our scripts also randomly sampled stack traces and error messages from their
co-occurring failures.
We ensured that the inspectors saw a reasonable diversity of stack traces by 
having our scripts take into account their pairwise Levenshtein distance when 
sampling.
The four inspectors then met and arrived at a collective answer to each question
for every cluster they were allocated following a process of negotiated 
agreement.
Negotiated agreement is a process whereby multiple researchers independently
analyze a dataset and then systematically compare their findings and discuss 
discrepancies until all researchers converge on a shared interpretation.
It features in the methodologies of previous software engineering 
studies~\cite{Hilton2017,Parry2022c}.
One inspector later reviewed the answers to the two questions regarding root 
causes and possible repairs for each cluster and identified general themes.

\subsection{Threats to Validity}
\label{sec:methodology:threats}

While we designed our methodology to rigorously investigate systemic flakiness, 
we must acknowledge several threats to validity.

{\bf Internal Validity.} Our clustering analysis relies on the Jaccard distance 
between sets of failing test suite runs.
While this metric effectively captures co-occurrence, other distance metrics 
might yield different clusters, potentially affecting our results.
Additionally, the automated selection of the clustering threshold based on the 
silhouette score introduces an inherent dependence on this specific metric.
While we set a minimum silhouette score of 0.6 to ensure well-defined clusters, 
borderline cases may have been excluded or misclassified. 
Errors in the implementation of our scripts could also impact results.
To mitigate this risk, we delegated the most important components to 
well-established and thoroughly tested third-party libraries such as SciPy and 
scikit-learn~\cite{Scipy2025,ScikitLearn2025}.

{\bf External Validity.} The generalizability of our findings is inherently 
limited by the dataset we used. 
While it represents a broad range of open-source Java projects from diverse 
domains, the dataset may not capture systemic flakiness behaviors in other 
programming languages. 
Moreover, it is possible that the 10,000 test suite runs were not sufficient to 
manifest all the flaky tests in the projects' test suites and to reliably 
observe the failure patterns of flaky tests with very low failure rates.
As pointed out by the original dataset authors~\cite{Alshammari2021} (and 
implied by other practitioners~\cite{Harman2018}), this is a general threat to 
the validity of any empirical study on flaky tests that cannot be totally 
rectified, but only mitigated by performing as many test suite runs as 
computationally feasible.

{\bf Construct Validity.} This study assumes that co-occurrence of flaky test 
failures is a symptom of systemic flakiness. 
While this assumption is well-founded, there may be other factors contributing 
to failure co-occurrence, including random chance. 
Additionally, our use of static test case distance measures as predictors of 
systemic flakiness is based on intuition and prior work on test suite 
diversity~\cite{Elgendy2024,Elgendy2025}.
However, these measures may not fully capture the complexity of relationships 
between flaky tests.

{\bf Conclusion Validity.} Our scripts calculated SHAP values to evaluate the 
importance of features in predicting systemic flakiness.
It is important to note that SHAP values only capture a feature's contribution 
to a model's prediction rather than a feature's value to the prediction task in 
general.
In other words, a feature with a high mean absolute SHAP value with respect to a
poorly performing model may not actually indicate anything about that feature's
predictive power.
The qualitative analysis for {\bf RQ3} is subject to potential bias due to its 
reliance on subjective interpretation. 
To mitigate this, we employed a process of negotiated agreement among the 
inspectors and randomly allocated clusters to ensure diverse perspectives. 
However, the findings may still reflect variations in individual expertise or 
interpretation.

    \section{Results}
\label{sec:results}

\subsection{Results for RQ1 (Prevalence)}
\label{sec:results:rqone}

\begin{table*}[t]
    \centering
    \caption{
        \label{tab:clusters}
        Results of the agglomerative clustering of flaky tests based on their
        failure co-occurrence.
        The table gives the number of flaky tests (Flaky Tests) and the 
        number of clusters (Clusters).
        For projects with at least one cluster, it also gives the number of 
        flaky tests that belong to a cluster (Flaky in Cluster), the mean number 
        of flaky tests per cluster (Mean Size), the mean number of test classes 
        per cluster (Mean Classes), the mean silhouette score (Silhouette), and 
        the distance threshold (Threshold).
    }
    \begin{tabular}{lrrrrrrr}
        \toprule
        {\bf Project Name} & 
        {\bf Flaky Tests} & 
        {\bf Clusters} & 
        {\bf Flaky in Cluster} & 
        {\bf Mean Size} &
        {\bf Mean Classes} & 
        {\bf Silhouette} & 
        {\bf Threshold} \\
        \midrule
        activiti-activiti & 32 & 0 & - & - & - & - & - \\
\rowcolor{gray!20}
Alluxio-alluxio & 116 & 1 & 113 & 113.00 & 16.00 & 0.88 & 0.52 \\
apache-ambari & 52 & 2 & 50 & 25.00 & 1.50 & 0.96 & 0.00 \\
\rowcolor{gray!20}
apache-hbase & 145 & 9 & 135 & 15.00 & 4.78 & 0.91 & 0.09 \\
apache-httpcore & 22 & 0 & - & - & - & - & - \\
\rowcolor{gray!20}
apache-incubator-dubbo & 19 & 0 & - & - & - & - & - \\
doanduyhai-Achilles & 4 & 0 & - & - & - & - & - \\
\rowcolor{gray!20}
elasticjob-elastic-job-lite & 3 & 1 & 2 & 2.00 & 1.00 & 0.67 & 0.00 \\
hector-client-hector & 33 & 1 & 31 & 31.00 & 7.00 & 0.94 & 0.00 \\
\rowcolor{gray!20}
jknack-handlebars.java & 1 & 0 & - & - & - & - & - \\
joel-costigliola-assertj-core & 1 & 0 & - & - & - & - & - \\
\rowcolor{gray!20}
kevinsawicki-http-request & 18 & 3 & 18 & 6.00 & 1.00 & 0.99 & 0.01 \\
ninjaframework-ninja & 1 & 0 & - & - & - & - & - \\
\rowcolor{gray!20}
orbit-orbit & 7 & 0 & - & - & - & - & - \\
qos-ch-logback & 22 & 0 & - & - & - & - & - \\
\rowcolor{gray!20}
spring-projects-spring-boot & 163 & 8 & 154 & 19.25 & 4.12 & 0.94 & 0.01 \\
square-okhttp & 100 & 10 & 74 & 7.40 & 1.10 & 0.74 & 0.00 \\
\rowcolor{gray!20}
tootallnate-java-websocket & 23 & 0 & - & - & - & - & - \\
undertow-io-undertow & 7 & 0 & - & - & - & - & - \\
\rowcolor{gray!20}
wildfly-wildfly & 23 & 6 & 18 & 3.00 & 1.00 & 0.78 & 0.00 \\
wro4j-wro4j & 16 & 4 & 11 & 2.75 & 2.00 & 0.64 & 0.11 \\
\rowcolor{gray!20}
zxing-zxing & 2 & 0 & - & - & - & - & - \\

\midrule
{\bf Overall} & 810 & 45 & 606 & 13.47 & 2.91 & - & - \\

        \bottomrule
    \end{tabular}
\end{table*}

Table~\ref{tab:clusters} shows the results of the agglomerative clustering of 
the flaky tests in the projects of the FlakeFlagger dataset based on the extent
of their failure co-occurrence.
The table gives the number of flaky tests and the number of clusters per project 
after extracting the concrete clustering with the greatest mean silhouette 
score.
For each project that contains at least one cluster, it also gives the number of 
flaky tests that belong to a cluster, the mean number of flaky tests per 
cluster, the mean number of distinct test classes per cluster, and finally the 
mean silhouette score and distance threshold.
The total run time required to compute the clusters for every project was 1.8 
seconds on a machine with an Intel Core i7-13700 CPU.

Of the 22 projects in the FlakeFlagger dataset that contain at least one flaky 
test, 10 (45\%) contain at least one cluster.
The remainder have either only a single flaky test or no possible concrete 
clustering with a mean silhouette score of at least 0.6.
There are 810 flaky tests and 45 clusters between the 22 projects.
Of the 810 flaky tests, 606 (75\%) belong to a cluster.
The mean number of flaky tests per cluster varies considerably between projects.
The mean size over the 45 clusters is 13.5 flaky tests.
On average, clusters contain flaky tests from 2.9 distinct test classes.
These results indicate that systemic flakiness is widespread and extends beyond 
test class boundaries.

\conclusionbox{RQ1}{%
    Systemic flakiness is a widespread and significant phenomenon.
    There are 45 clusters between the 22 projects in the FlakeFlagger dataset 
    that contain flaky tests.
    Of the 810 flaky tests in the dataset, 606 (75\%) belong to a cluster.
    The mean cluster size is 13.5 flaky tests.
}

\subsection{Results for RQ2 (Prediction)}
\label{sec:results:rqtwo}

\begin{table*}[t]
    \centering
    \caption{
        \label{tab:estimation}
        The performance of three machine learning models for predicting systemic 
        flakiness using static test case distance measures in projects with at 
        least 10 flaky tests and one cluster.
        The table gives the performance for the regression task using the 
        coefficient of determination (\rsq) and the classification task using 
        the Matthews correlation coefficient (MCC).
    }
    \begin{tabular}{lrrrrrr}
        \toprule
        & 
        \multicolumn{2}{c}{\bf Extra Trees} & 
        \multicolumn{2}{c}{\bf Gradient Boosting} & 
        \multicolumn{2}{c}{\bf Random Forest} \\
        \cmidrule(r){2-3}
        \cmidrule(rl){4-5}
        \cmidrule(r){6-7}
        &
        {\bf Regression} &
        {\bf Classification} &
        {\bf Regression} &
        {\bf Classification} &
        {\bf Regression} &
        {\bf Classification} \\
        {\bf Project Name} &
        {\bf (\rsq)} &
        {\bf (MCC)} &
        {\bf (\rsq)} &
        {\bf (MCC)} &
        {\bf (\rsq)} &
        {\bf (MCC)} \\
        \midrule
        Alluxio-alluxio & 0.51 & 0.41 & 0.27 & 0.24 & 0.48 & 0.39 \\
\rowcolor{gray!20}
apache-ambari & 0.98 & 0.99 & 0.97 & 0.98 & 0.96 & 0.97 \\
apache-hbase & 0.80 & 0.84 & 0.63 & 0.74 & 0.77 & 0.83 \\
\rowcolor{gray!20}
hector-client-hector & 0.87 & 0.95 & 0.76 & 0.85 & 0.78 & 0.88 \\
kevinsawicki-http-request & 0.88 & 0.77 & 0.85 & 0.65 & 0.82 & 0.66 \\
\rowcolor{gray!20}
spring-projects-spring-boot & 0.95 & 0.96 & 0.87 & 0.92 & 0.93 & 0.96 \\
square-okhttp & 0.36 & 0.29 & 0.37 & 0.19 & 0.38 & 0.27 \\
\rowcolor{gray!20}
wildfly-wildfly & 0.92 & 0.87 & 0.86 & 0.87 & 0.88 & 0.75 \\
wro4j-wro4j & 0.37 & 0.62 & 0.36 & 0.23 & 0.38 & 0.35 \\
\midrule
{\bf Mean} & 0.74 & 0.74 & 0.66 & 0.63 & 0.71 & 0.67 \\

        \bottomrule
    \end{tabular}
\end{table*}

\begin{table*}[t]
    \centering
    \caption{
        \label{tab:importance}
        The ranks of the 21 static test case distances measures in terms of 
        their mean absolute SHAP value with respect to the regression task and 
        the extra trees model.
        Lower ranks indicate that the distance measure has a greater mean 
        absolute SHAP values and was thus a more important feature.
    }
    \begin{tabular}{llrrrrrrrrrr}
        \toprule
        & & \multicolumn{10}{c}{\bf Importance Rank} \\
        \cmidrule{3-12}
        {\bf Distance} & 
        {\bf Applied to} &
        \rotatebox{77}{\bf Alluxio-alluxio} &
        \rotatebox{77}{\bf apache-ambari} &
        \rotatebox{77}{\bf apache-hbase} &
        \rotatebox{77}{\bf hector-client...} &
        \rotatebox{77}{\bf kevinsawicki-http...} &
        \rotatebox{77}{\bf spring-projects...} &
        \rotatebox{77}{\bf square-okhttp} &
        \rotatebox{77}{\bf wildfly-wildfly} &
        \rotatebox{77}{\bf wro4j-wro4j} &
        {\bf Mean} \\
        \midrule
        Hierarchy & Name & 1 & 1 & 1 & 9 & 21 & 1 & 1 & 1 & 3 & 4.33 \\
\rowcolor{gray!20}
Overlap & Name & 9 & 2 & 2 & 10 & 4 & 6 & 10 & 18 & 2 & 7.00 \\
Levenshtein & Name & 3 & 16 & 14 & 5 & 11 & 9 & 2 & 8 & 12 & 8.89 \\
\rowcolor{gray!20}
Normalized Levenshtein & Name & 13 & 3 & 3 & 19 & 15 & 7 & 9 & 3 & 8 & 8.89 \\
Overlap & Code & 6 & 10 & 15 & 3 & 1 & 2 & 14 & 10 & 20 & 9.00 \\
\rowcolor{gray!20}
Damerau-Levenshtein & Name & 2 & 14 & 16 & 7 & 12 & 8 & 3 & 6 & 14 & 9.11 \\
Dice & Name & 7 & 5 & 9 & 17 & 14 & 11 & 6 & 9 & 5 & 9.22 \\
\rowcolor{gray!20}
Normalized Damerau-Levenshtein & Name & 18 & 4 & 4 & 16 & 10 & 13 & 5 & 7 & 7 & 9.33 \\
Jaro & Name & 15 & 6 & 7 & 15 & 17 & 12 & 11 & 2 & 1 & 9.56 \\
\rowcolor{gray!20}
Normalized Compression & Name & 11 & 9 & 5 & 11 & 16 & 18 & 4 & 5 & 9 & 9.78 \\
Jaccard & Name & 8 & 7 & 12 & 18 & 6 & 10 & 13 & 11 & 6 & 10.11 \\
\rowcolor{gray!20}
Levenshtein & Code & 5 & 12 & 8 & 14 & 2 & 14 & 7 & 12 & 21 & 10.56 \\
Damerau-Levenshtein & Code & 4 & 11 & 10 & 13 & 3 & 16 & 8 & 13 & 19 & 10.78 \\
\rowcolor{gray!20}
Dice & Code & 17 & 13 & 13 & 1 & 7 & 3 & 15 & 21 & 16 & 11.78 \\
Jaccard & Code & 16 & 15 & 17 & 2 & 5 & 4 & 16 & 20 & 15 & 12.22 \\
\rowcolor{gray!20}
Jaro-Winkler & Name & 20 & 8 & 11 & 20 & 18 & 19 & 12 & 4 & 4 & 12.89 \\
Normalized Compression & Code & 10 & 17 & 6 & 21 & 13 & 5 & 17 & 16 & 13 & 13.11 \\
\rowcolor{gray!20}
Normalized Levenshtein & Code & 12 & 18 & 19 & 8 & 8 & 17 & 18 & 17 & 18 & 15.00 \\
Normalized Damerau-Levenshtein & Code & 14 & 21 & 18 & 12 & 9 & 15 & 19 & 19 & 17 & 16.00 \\
\rowcolor{gray!20}
Jaro & Code & 19 & 20 & 20 & 4 & 20 & 20 & 21 & 15 & 11 & 16.67 \\
Jaro-Winkler & Code & 21 & 19 & 21 & 6 & 19 & 21 & 20 & 14 & 10 & 16.78 \\

        \bottomrule
    \end{tabular}
\end{table*}

Table~\ref{tab:estimation} presents the performance of three machine learning 
models at using static test case distance measures to predict systemic flakiness
in projects with at least 10 flaky tests and one cluster.
For each type of model (extra trees, gradient boosting, and random forest), the 
table gives the performance for the regression task using the coefficient of 
determination (\rsq) and the classification task using the Matthews 
correlation coefficient (MCC).
For each pair of flaky tests in a project, the regression task is to predict the 
Jaccard distance between the two sets of failing run IDs and the classification
task is to predict whether they belong to the same cluster.
The total run time required to train and evaluate every model for each project 
for both tasks was 210 seconds with an Intel Core i7-13700 CPU.

For the regression task, the extra trees model has the greatest mean performance 
in terms of \rsq at 0.74.
This means that, on average, 74\% of the variance in the Jaccard distance over
every pair of flaky tests is explained by the model using the static test case 
distance measures described in Section~\ref{sec:methodology:rqtwo} as features.
There is inconsistency in the per-project performance of all three models.
Focusing on the extra trees model, the performance is reasonable for most 
projects, achieving an \rsq of or above 0.8 for 6 out of 9 projects.
It is particularly good for apache-ambari, spring-projects-spring-boot, and 
wildfly-wildfly, where it achieves an \rsq above 0.9.
The performance is quite poor for the projects square-okhttp and wro4j-wro4j 
where the model fails to explain even half of the variance.
This pattern is reflected by the other two models to varying extents, indicating 
that it is due to properties of these projects and the features rather than the 
extra trees model specifically.

For the classification task, the extra trees model has the greatest mean 
performance in terms of MCC at 0.74.
The per-project performance for this task appears roughly correlated with that 
for the regression task.
This is unsurprising, given they both use the same features and evaluate the 
ability to predict systemic flakiness.

Table~\ref{tab:importance} gives the ranks of each static test case distance
measure in terms of their mean absolute SHAP value with respect to the 
regression task and extra trees model per project.
A rank of 1 indicates that the distance measure has the greatest mean absolute 
SHAP value and was thus the feature with the greatest impact on the model's 
prediction.
A rank of 21 indicates the opposite.
The table also gives the mean rank for each feature over every project.
According to the mean ranks, the hierarchy distance was the most important 
feature and distance measures applied to the names of test cases were generally
more important than those applied to the source code.
However, the per-project ranks for each feature vary significantly.
In the case of the hierarchy distance, it was the least important feature for
the kevinsawicki-http-request project despite being the most important feature 
on average.
\pagebreak
\conclusionbox{RQ2}{%
    Machine learning models can predict systemic flakiness using static test 
    case distance measures to varying extents across projects.
    The extra trees model has the greatest mean performance at the regression 
    task, achieving an \rsq of 0.74, and the classification task, achieving an
    MCC of 0.74.
    On average, the hierarchy distance is the most important feature and 
    distance measures applied to the names of test cases are generally more 
    important than those applied to the code.
}

\subsection{Results for RQ3 (Causes)}
\label{sec:results:rqthree}

\begin{table}[t]
    \centering
    \caption{
        \label{tab:causes}
        Cause theme frequencies from our qualitative analysis of the 45 
        clusters.
        The totals sum to greater than 45 because some clusters belong to 
        multiple themes.
    }
    \begin{tabular}{lrrrrrr}
        \toprule
        {\bf Project Name} & 
        \rotatebox{90}{\bf Networking} & 
        \rotatebox{90}{\bf External Dependency} & 
        \rotatebox{90}{\bf Filesystem Pollution} & 
        \rotatebox{90}{\bf Timeout} &
        \rotatebox{90}{\bf Unknown} &
        \rotatebox{90}{\bf System Clock} \\
        \midrule
        Alluxio-alluxio & 1 & 0 & 0 & 0 & 0 & 0 \\
\rowcolor{gray!20}
apache-ambari & 2 & 0 & 0 & 0 & 0 & 0 \\
apache-hbase & 6 & 0 & 5 & 3 & 0 & 0 \\
\rowcolor{gray!20}
elasticjob-elastic-job-lite & 0 & 0 & 0 & 0 & 0 & 1 \\
hector-client-hector & 0 & 1 & 0 & 0 & 0 & 0 \\
\rowcolor{gray!20}
kevinsawicki-http-request & 3 & 0 & 0 & 0 & 0 & 0 \\
spring-projects-spring-boot & 1 & 6 & 0 & 0 & 1 & 0 \\
\rowcolor{gray!20}
square-okhttp & 10 & 0 & 0 & 0 & 0 & 0 \\
wildfly-wildfly & 1 & 5 & 0 & 1 & 0 & 0 \\
\rowcolor{gray!20}
wro4j-wro4j & 1 & 2 & 0 & 0 & 1 & 0 \\
\midrule
{\bf Total} & 25 & 14 & 5 & 4 & 2 & 1 \\

    \bottomrule
    \end{tabular}
\end{table}

\begin{figure*}
    \begin{lstlisting}[language=Java,basicstyle=\footnotesize]
@Test(expected = BadCredentialsException.class)
public void testBadCredential() throws Exception {
    Authentication authentication = new UsernamePasswordAuthenticationToken("notFound", "wrong");
    authenticationProvider.authenticate(authentication);
}
@Test
public void testAuthenticate() throws Exception {
    assertNull("User alread exists in DB", userDAO.findLdapUserByName("allowedUser"));
    Authentication authentication = new UsernamePasswordAuthenticationToken("allowedUser", "password");
    Authentication result = authenticationProvider.authenticate(authentication);
    assertTrue(result.isAuthenticated());
    assertNotNull("User was not created", userDAO.findLdapUserByName("allowedUser"));
    result = authenticationProvider.authenticate(authentication);
    assertTrue(result.isAuthenticated());
}
    \end{lstlisting}
    \caption{
        \label{fig:networking}
        Two flaky tests from the apache-ambari project that form a {\it 
        Networking} cluster.
        They both failed after calling the {\tt authenticate} method during the 
        exact same 8 test suite runs out of 10,000.
        In both cases, the root exception was {\tt java.net.ConnectException: 
        Connection refused (Connection refused)}.
    }
\end{figure*}

Table~\ref{tab:causes} summarizes the results of our qualitative analysis of the 
causes of the 45 clusters in the FlakeFlagger dataset.
It gives the frequencies of each cause theme that we identified.
Clusters for which we were unable to identify the cause are represented by the
{\it Unknown} theme.
The totals sum to greater than the number of clusters because some clusters
belong to multiple themes.

The most common theme is {\it Networking}.
This represents clusters where an intermittent networking issue causes the 
failure of a group of flaky tests during a single test suite run.
In these instances, the test case tries to establish a connection over the 
network or executes production code that does.
Examples of such networking issues include DNS resolution failures, an 
unreachable network, and timeouts due to the network taking longer than expected 
to process requests.
Generally speaking, we found clusters in this category to be the largest and 
most diverse in terms of their member flaky tests.
This is probably because an intermittent networking issue is likely to impact 
any test case that requires a functional network connection regardless of its
purpose or which part of the production code that it tests.
See Figure~\ref{fig:networking} for an example of a {\it Networking} cluster.
Please refer to the replication package~\cite{ReplicationPackage} for more 
examples.

The second most common theme is {\it External Dependency}.
Clusters of this theme contain groups of flaky tests that all depend on some
external service, library, or other artifact that is outside the control of the
software under test.
If the external service is unavailable or is exhibiting unexpected behavior 
during a particular test suite run, then all the flaky tests that depend upon it
are likely to fail at the same time.
In one cluster of the spring-projects-spring-boot project, the external 
dependency issue appeared to be related to non-determinism in the version of 
Spring Framework that was being installed when building the project before each
test suite run.
In another cluster of the wildfly-wildfly project, flaky tests were 
reliant on an external web server that was intermittently returning 500 errors,
which is an HTTP status code indicating a server-side error.
We did not consider this to be an instance of the {\it Networking} theme because 
the server encountered some sort of intermittent problem and the network itself 
did not fail.

Less common themes are {\it Filesystem Pollution}, {\it Timeout}, and {\it 
System Clock}.
In {\it Filesystem Pollution}, clusters are caused by one or more test cases 
modifying the filesystem (\eg, creating a directory) and then failing in such a 
way that they omit to perform proper clean up procedures.
This triggers the failure of a group of subsequent test cases due to the 
unexpected state of the filesystem.
In {\it Timeout}, a hard coded time limit for some event to occur triggers the 
failure of a group of flaky tests.
This typically co-occurs with the {\it Networking} theme where the time limit 
concerns a response from a server.
We observed only a single cluster under the {\it System Clock} theme, where a 
small cluster of flaky tests that made direct use of the system clock failed at 
the same time.

\conclusionbox{RQ3}{%
    Following qualitative analysis of the causes of the 45 clusters, the most 
    common theme is {\it Networking}.
    This represents clusters where an intermittent networking issue causes the 
    failure of a group of flaky tests during a single test suite run.
    The second most common theme is {\it External Dependency}.
    Clusters of this theme contain groups of flaky tests that all depend on some 
    external service, library, or other artifact that is outside the control of 
    the software under test.
}

    \section{Discussion}
\label{sec:discussion}

\subsection{Distance Thresholds}
\label{sec:discussion:distance}

\begin{figure*}[t]
    \centering
    \begin{subfigure}{0.33\textwidth}
        \centering
        \includegraphics[width=\textwidth]{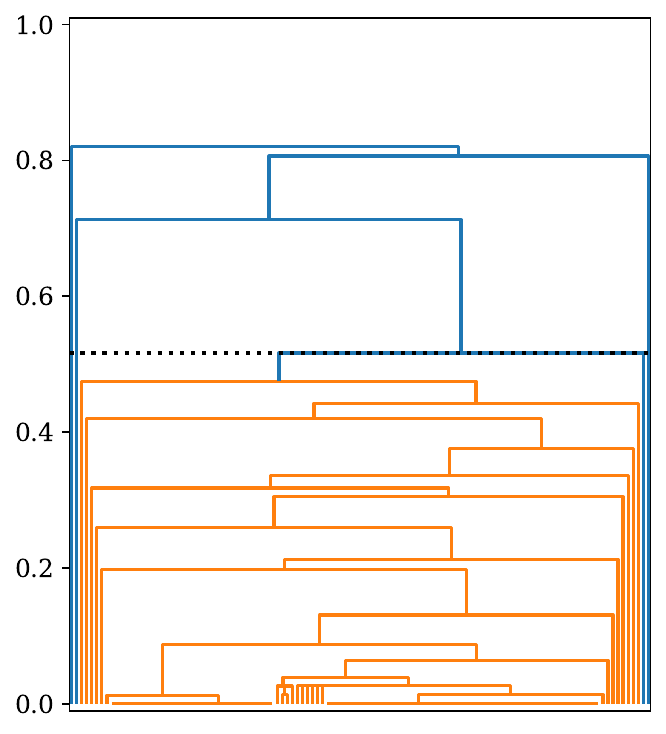}
        \caption{
            \label{fig:dendrograms:alluxio}
            Alluxio-alluxio
        }
    \end{subfigure}
    \begin{subfigure}{0.33\textwidth}
        \centering
        \includegraphics[width=\textwidth]{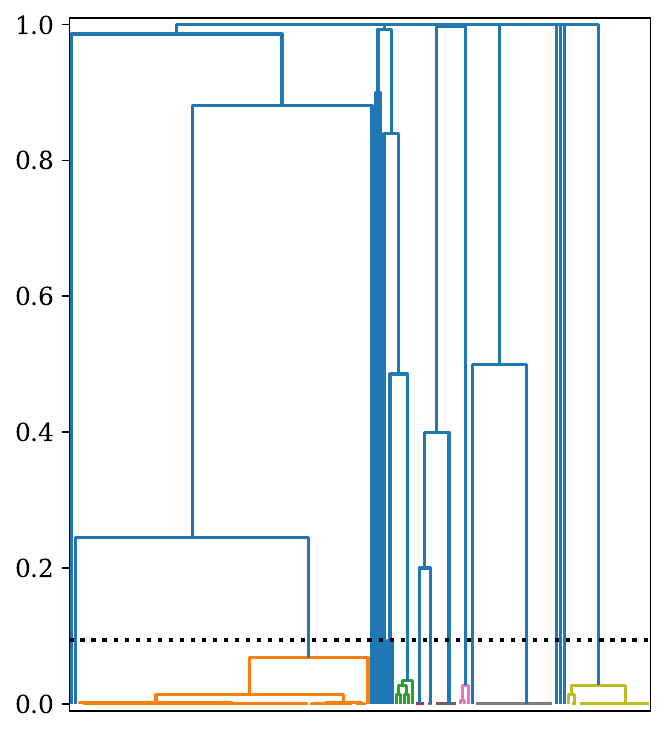}
        \caption{
            \label{fig:dendrograms:apache}
            apache-hbase
        }
    \end{subfigure}
    \begin{subfigure}{0.33\textwidth}
        \centering
        \includegraphics[width=\textwidth]{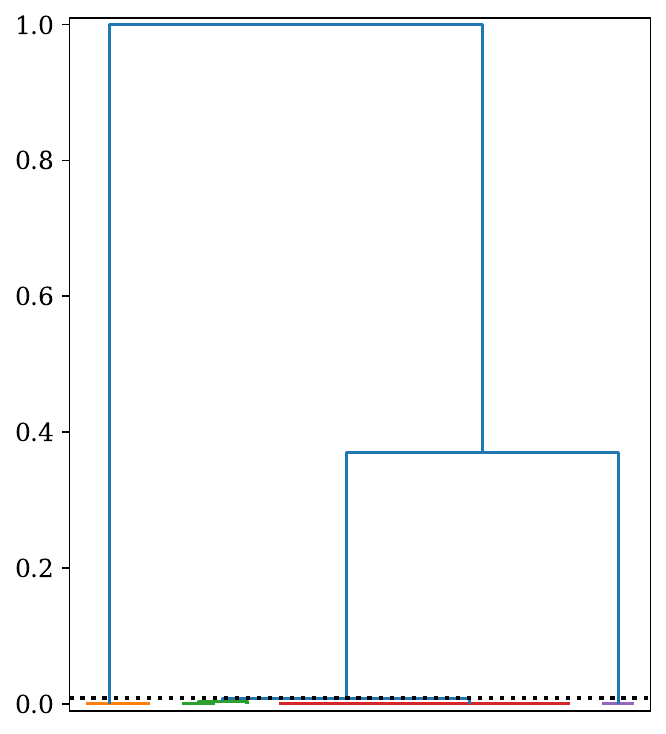}
        \caption{
            \label{fig:dendrograms:kevinsawicki}
            kevinsawicki-http-request
        }
    \end{subfigure}
    \caption{
        \label{fig:dendrograms}
        Dendrograms for three projects illustrating the hierarchy of the 
        clusters prior to extracting a concrete clustering.
        The vertical axis shows the Jaccard distance at which clusters are 
        merged (see Equation~\ref{equ:jaccard}).
        The dotted line represents the distance threshold that produces the 
        concrete clustering with the greatest mean silhouette score (see 
        Equation~\ref{equ:silhouette}).
    }
\end{figure*}

Figure~\ref{fig:dendrograms} illustrates the hierarchy of the clusters, prior 
to extracting a concrete clustering, in the form of a dendrogram for three 
projects.
Each dendrogram represents a tree structure, where leaves represent flaky tests 
and branches represent how they are progressively clustered based on their 
failure co-occurrence.
The vertical axis shows the Jaccard distance at which clusters are merged (see 
Equation~\ref{equ:jaccard}), with lower branches representing more similar sets
of failing run IDs between flaky tests.
The dotted line represents the distance threshold that produces the concrete
clustering with the greatest mean silhouette score (see 
Equation~\ref{equ:silhouette}), as identified by our scripts.
The clusters are color coded, with blue representing individual flaky tests that 
did not make it into a non-singleton cluster.
Other clusters are assigned distinct colors to visually differentiate them, 
though the specific colors do not carry inherent meaning.

For Alluxio-alluxio, there are many branches distributed fairly evenly as the
distance increases up to about 0.8.
This indicates that the sets of failing run IDs of the flaky tests in this 
project are rather diverse.
There is a single cluster for this project and the distance threshold is 0.52, 
which indicates a cluster of flaky tests with relatively heterogeneous (but 
still related) sets of failing run IDs.
For kevinsawicki-http-request, the branches are mainly concentrated at the very
bottom of the dendrogram.
This indicates flaky tests with very similar sets of failing run IDs.
There are three clusters for this project and the distance threshold is 0.01, 
which indicates clusters of flaky tests with almost identical sets of failing 
run IDs.
The dendrogram for apache-hbase represents a situation somewhere in between.
The threshold for five projects is 0, indicating clusters of flaky tests with 
strictly identical sets of failing run IDs.

\subsection{Repairs and Mitigations}
\label{sec:discussion:repairs}

\begin{table}[t]
    \centering
    \caption{
        \label{tab:repairs}
        Repair/mitigation theme frequencies from our qualitative analysis of the 
        45 clusters.
        The totals sum to greater than 45 because some clusters belong to 
        multiple themes.
    }
    \begin{tabular}{lrrrrrr}
        \toprule
        {\bf Project Name} & 
        \rotatebox{90}{\bf Look Before You Leap} & 
        \rotatebox{90}{\bf Unknown} &
        \rotatebox{90}{\bf Mocking} & 
        \rotatebox{90}{\bf Better Setup/Teardown} & 
        \rotatebox{90}{\bf Better Error Checking} & 
        \rotatebox{90}{\bf Avoid Networking} \\
        \midrule
        Alluxio-alluxio & 1 & 0 & 1 & 0 & 0 & 0 \\
\rowcolor{gray!20}
apache-ambari & 1 & 0 & 0 & 2 & 0 & 0 \\
apache-hbase & 1 & 1 & 4 & 2 & 3 & 0 \\
\rowcolor{gray!20}
elasticjob-elastic-job-lite & 0 & 0 & 1 & 0 & 0 & 0 \\
hector-client-hector & 0 & 1 & 0 & 0 & 0 & 0 \\
\rowcolor{gray!20}
kevinsawicki-http-request & 1 & 0 & 0 & 0 & 0 & 3 \\
spring-projects-spring-boot & 1 & 7 & 1 & 0 & 0 & 0 \\
\rowcolor{gray!20}
square-okhttp & 10 & 0 & 0 & 0 & 0 & 0 \\
wildfly-wildfly & 5 & 1 & 0 & 0 & 0 & 0 \\
\rowcolor{gray!20}
wro4j-wro4j & 0 & 4 & 0 & 0 & 0 & 0 \\
\midrule
{\bf Total} & 20 & 14 & 7 & 4 & 3 & 3 \\

    \bottomrule
    \end{tabular}
\end{table}

One of the questions the inspectors answered about each cluster during the
qualitative analysis was ``What actions could a developer take to repair or 
mitigate the flaky tests in this cluster?''.
Table \ref{tab:repairs} gives the frequencies of the general themes in the
collective answers to this question.
Generally speaking, it was difficult for the inspectors to confidently answer 
this question because they were not familiar with the intricacies of the 
projects involved.

The most common theme by far was {\it Look Before You Leap}.
This represents clusters of flaky tests that could be repaired or at least 
mitigated by checking the status of some external system or resource.
Examples include checking for the existence of a directory and confirming that
a server is running.
The second most common theme was {\it Mocking}.
Clusters of this theme could be addressed through proper mocking of some third 
party library or external service~\cite{Spadini2017}.
For example, the cluster of the {\it System Clock} cause theme could have 
been mitigated by mocking the system clock.

In {\it Better Setup/Teardown}, inspectors identified possible improvements to 
the setup and teardown methods that are executed before and after test cases.
In a recent developer survey, respondents rated issues in setup and teardown
methods to be the most common causes of flaky tests~\cite{Parry2022b}.
In {\it Better Error Checking}, clusters of flaky tests failed with nondescript
error messages such as {\tt NullPointerException}, caused by intermittent
problems deep in the call stack.
Inspectors suggested more comprehensive error checking to facilitate debugging,
making this theme more of a mitigation strategy than a bona fide repair.
Finally, in {\it Avoid Networking}, inspectors determined that the underling
program logic being evaluated by the flaky tests could have tested independently
of any networking, which was the cause of the flakiness.

\subsection{Implications and Future Directions}
\label{sec:discussion:implications}

This study is the first to characterize systemic flakiness and represents an
inflection point in flaky test research.
The findings have major implications for developers and researchers, and open up 
a myriad of avenues for future work.

{\bf Cost and Developer Impact.} This study found that the mean cluster size 
over all projects was 13.5 flaky tests.
An industrial case study on the cost of flaky tests in continuous integration 
found that developers allocated up to 1.28\% of their time to repairing flaky 
tests, translating to a monthly cost of \$2,250~\cite{Leinen2024}.
By recognizing systemic flakiness, developers can achieve significant cost and 
time savings by resolving underlying root causes that simultaneously fix 
multiple flaky tests, rather than inefficiently debugging and repairing them in 
isolation.
Future studies should focus on developing automated techniques to detect and 
triage systemic flakiness in continuous integration pipelines to reduce the cost 
of flakiness.

{\bf Impact on Testing Techniques.} This study found that 75\% of flaky tests 
across all projects fail as part of a cluster.
Prior studies evaluated the negative impacts of simulated flakiness on fault 
localization, mutation testing, and automated program 
repair~\cite{Vancsics2020,Cordy2022}.
These studies did not consider systemic flakiness and did not simulate clusters 
of flaky tests with co-occurring failures.
Therefore, the impact of flaky tests on these techniques may be misrepresented.
Future studies should revisit this assessment using a more realistic simulation 
model that accounts for systemic flakiness.

{\bf Machine Learning for Systemic Flakiness Prediction.} This study found that 
machine learning models can predict systemic flakiness using static test case 
distance measures.
Previous studies have evaluated machine learning-based techniques to classify 
individual test cases as flaky or not~\cite{Pinto2020,Alshammari2021,
Bertolino2021,Camara2021,Parry2022,Pontillo2022,Fatima2022,Qin2022,Gruber2023b,
Parry2023}.
By not considering systemic flakiness, these techniques do not benefit from 
valuable contextual features, such as historical patterns of failure 
co-occurrence.
Without this information, techniques cannot identify inter-test relationships, 
leading to predictions that focus on isolated flaky tests rather than underlying 
root causes.
This may limit a technique's ability to generalize across projects, which is a 
well-known limitation~\cite{Afeltra2024}.
Future studies should explore integrating systemic flakiness prediction into 
existing flaky test classification techniques and evaluate its impact on 
cross-project generalization.

{\bf Feature Engineering.} This study found that the hierarchy distance was the 
most important feature for systemic flakiness prediction, and distance measures 
applied to the names of test cases were generally more important than those 
applied to the source code.
Prior studies have established that the inclusion of dynamic features enhances 
machine learning-based flaky test detection~\cite{Alshammari2021,Parry2022}.
Dynamic test case distance measures, such as the Jaccard distance between 
coverage profiles, may capture additional signals that static test case distance 
measures miss.
Future studies should refine the feature set and explore additional metrics.

{\bf Causes and Mitigation Strategies.} This study identified intermittent 
networking issues and instabilities in external dependencies as predominant 
causes of systemic flakiness through manual 
inspection.
In contrast, previous studies that categorized the causes of individual flaky 
tests generally rated asynchronous operations and concurrency as the leading 
causes~\cite{Luo2014b,Vahabzadeh2015,Eck2019,Lam2020c,Gruber2021,Hashemi2022}.
By overlooking systemic flakiness, these studies may have unintentionally 
presented a skewed distribution of flaky test causes.
Future studies should conduct larger-scale empirical analyses of the causes of 
systemic flakiness across different programming languages and testing 
frameworks to rectify this.
This study also found that clusters contain flaky tests from 2.9 distinct test 
classes on average.
This indicates that developers should not only take greater care to isolate test 
cases from environmental variability but should also ensure that test classes 
are properly decoupled.
Future studies should focus on automated techniques to assist developers in this 
regard.

{\bf Automating Root Cause Analysis.} This study involved manual inspection of 
each cluster, which was a time-consuming process and in some cases did not 
identify any causes.
Future studies should leverage large language models to automate the 
identification of systemic flakiness causes by analyzing test code and stack 
traces.

{\bf Developer Perception.} This study was based on quantitative analysis of 
test execution data and manual inspection conducted by the authors.
Industrial software developers were not directly involved in the methodology.
Understanding how developers currently perceive and address systemic flakiness 
is crucial for designing techniques to address it.
Future studies should conduct developer surveys and interviews to assess whether 
developers are aware of systemic flakiness and, if so, how they deal with it.

{\bf Benchmarking and Dataset Creation.} This study relied on an established 
dataset of flaky test failures, but existing datasets do not explicitly capture 
systemic flakiness.
Future studies should focus on creating benchmark datasets that annotate failure 
co-occurrence, enabling further research into systemic flakiness detection and 
mitigation.
It would be beneficial for the purposes of generalizability if projects written
in multiple programming languages were represented.
These datasets could be used to evaluate new machine learning models and 
continuous integration strategies.

    \section{Related Work}
\label{sec:related}

Golagha \etal proposed a technique to cluster failing hardware-in-the-loop tests 
based on non-code-based features in the absence of coverage data, which they 
argued is difficult to acquire in this domain~\cite{Golagha2019}.
Both their study and this study grouped failing test cases using agglomerative
clustering, but with different aims.
The aim in their study was to reduce manual debugging effort while still 
identifying as many bugs as possible by selecting only a single representative 
failing test case from each cluster for developers to review. 
The aim in this study was to identify instances of systemic flakiness by 
clustering flaky tests based on failure co-occurrence.

An \etal proposed a machine learning-based technique to automatically identify 
if a pair of failing test cases share the same root cause in the continuous 
integration pipeline of SAP HANA~\cite{An2022}.
Both their study and this study evaluated the ability of machine learning models 
to predict whether test failures share a common underlying cause based on 
pairwise similarity/distance measures, but with different aims.
The aim in their study was to reduce redundant bug reports and manual debugging 
effort.
The aim in this study was to evaluate the feasibility of a lightweight 
alternative to performing 10,000 test suite runs to identify systemic flakiness.

Prior studies on flaky test detection have explored machine learning techniques 
to identify flaky tests without requiring thousands of reruns~\cite{Pinto2020,
Alshammari2021,Bertolino2021,Camara2021,Parry2022,Pontillo2022,Fatima2022,
Qin2022,Gruber2023b,Parry2023}. 
Pinto \etal investigated whether flaky tests have a distinct ``vocabulary'' of 
identifiers and keywords, training machine learning classifiers on 
vocabulary-based features extracted from test case bodies~\cite{Pinto2020}. 
Their features included occurrences of whole identifiers and their components, 
along with complexity metrics such as lines of code. 
Alshammari \etal developed and evaluated FlakeFlagger, selecting 16 test case 
features as potential indicators of flakiness. 
These included eight boolean features capturing test smells~\cite{Garousi2018} 
and several numeric features, such as lines of code, number of assertions, and 
production code coverage.
This study differs from these prior studies in that we apply machine learning to 
predict systemic flakiness, using static test case distance measures to estimate 
the Jaccard distance between the sets of failing run IDs of pairs of flaky 
tests.

Several previous studies have categorized flaky tests by their root causes via
manual inspection~\cite{Luo2014b,Vahabzadeh2015,Eck2019,Lam2020c,Gruber2021,
Hashemi2022}.
Luo \etal categorized 201 commits that repaired flaky tests from 51 
projects of varying size and language~\cite{Luo2014b}.
They identified asynchronous calls, concurrency bugs, and test order 
dependencies as the most common causes.
Eck \etal asked 21 software developers to categorize 200 flaky tests that they
had previously repaired~\cite{Eck2019}.
They also identified concurrency bugs and asynchronous calls as the most common
causes, corroborating the findings of Luo \etal, alongside overly restrictive
assertion ranges.
While these previous studies focus on the causes of individual flaky tests, this 
study examines clusters of flaky tests, revealing systemic causes such as 
networking issues and instabilities in external dependencies.

A significant body of work has addressed {\it order-dependent} flaky tests,
whose outcome depends on the execution order of test 
cases~\cite{Lam2019,Shi2019,Li2022,Li2022b,Wang2022}.
This is typically caused by side effects left behind by previously executed test 
cases in the global program state (\eg, static fields in Java) or in the 
filesystem.
This study highlighted filesystem pollution as one of the possible causes of
systemic flakiness.
However, reordering the test cases, which prior studies typically perform to
identify order-dependent flaky tests~\cite{Zhang2014,Lam2019}, was never part of 
the methodology of this study or of the study that produced the FlakeFlagger 
dataset~\cite{Alshammari2021}.
Therefore, systemic flakiness is a concept it is clearly distinct from 
order-dependent flakiness.

    \section{Conclusion}
\label{sec:conclusion}

This study established systemic flakiness as a widespread and significant 
phenomenon. 
Through agglomerative clustering, we found that 75\% of flaky tests in the 
dataset belong to a cluster, indicating that flaky tests frequently fail 
together.
We demonstrated that machine learning models can predict systemic flakiness 
using static test case distance measures.
Extra trees was the best performing model on average, achieving an \rsq of 0.74 
when predicting the Jaccard distance between the sets of failing run IDs of 
pairs of flaky tests.
The hierarchy distance measure was the most important feature on average in
terms of mean absolute SHAP value.
Manual inspection of flaky test clusters revealed that systemic flakiness is 
primarily driven by intermittent networking issues and instabilities in external 
dependencies.
These results emphasize that flaky tests often share causes that transcend 
individual test logic.

The prevalence of systemic flakiness has important implications for developers
because it shows that they can simultaneously repair multiple flaky tests by
addressing the underlying shared root causes.
It also has important implications for research because it challenges the 
assumption that flaky test failures are isolated occurrences.

As future work, we plan on extending our empirical study by evaluating a larger
set of projects from multiple programming languages.
In doing so, we will be able to assess the generalizability of our findings 
beyond Java projects.
This will also result in a comprehensive dataset specifically for studying 
systemic flakiness.

    \balance
    \bibliography{bibliography/bibliography,local_bibliography}
\end{document}